\documentstyle[aps,prl,twocolumn,epsfig]{revtex}
\begin{document}

\title{ Low-momentum Pion Enhancement Induced by Chiral Symmetry Restoration }
\author{Pengfei Zhuang\\
        Physics Department, Tsinghua University, Beijing 100084, China }
\maketitle

\begin{abstract}
The thermal and nonthermal pion production by sigma decay and its
relation with chiral symmetry restoration in a hot and dense
matter are investigated. The nonthermal decay into pions of sigma
mesons which are popularly produced in chiral symmetric phase
leads to a low-momentum pion enhancement as a possible signature
of chiral phase transition at finite temperature and density.
\end{abstract}
\noindent ${\bf Keywords: chiral\ symmetry,\ pion\ spectra}$
\noindent ${\bf PACS: 11.30.Rd, 14.40.-n}$\\

As is well known, the ultimate goal of relativistic heavy ion
collisions\cite{qm01} is to study the deconfinement process in
moving from a hadron gas to a quark-gluon plasma and the  chiral
transition from the chiral symmetry breaking phase to the phase in
which it is restored\cite{hwa}. In recent years some signatures of
the chiral transition have been proposed, such as the excess of
pions due to a rapid thermal and chemical equilibration in the
symmetric phase\cite{song}, the excess of low energy photon pairs
by pion annihilation in hot and dense medium\cite{volkov}, the
dilepton enhancement from leptonic decay of sigma\cite{weldon},
and the enhancement of the continuum threshold in the scalar
channel\cite{chiku}. Most of the signatures are associated with
the changes of sigma properties at finite temperatures and
densities. While in the vacuum there is no obvious reason for the
presence of sigma mesons in the study of chiral properties, there
is a definite need\cite{rho} for sigma at finite temperatures and
densities in the investigation of chiral symmetry restoration.
With increasing temperature and density sigma changes its
character from a resonance with large mass to a bound state with
small mass. This medium dependence of sigma mass makes important
sense not only in thermodynamics\cite{zhuang}, but also in
particle yields. Around the critical point of the chiral
transition sigma is one of the most numerous species, since it is
nearly massless. If the transition happens at the freeze-out these
light sigmas will lead to large event-by-event
fluctuations\cite{rapp} of final state pions due to the
$\sigma$-exchange in the pion thermodynamic
potential\cite{stephanov}.

A direct signature of these light sigmas in hot and dense matter
is a nonthermal excess\cite{stephanov} of pions due to the decay
process $\sigma\rightarrow 2\pi$. With the expansion of the
hadronic system produced in relativistic heavy ion collisions, the
in-medium sigma mass rises towards its vacuum value and eventually
exceeds the $\pi\pi$ threshold. As the $\sigma\pi\pi$ coupling is
large, the decay proceeds rapidly. When this process occurs before
freeze-out, the produced pions will be thermalized in the heat
bath. However, when the decay happens after freeze-out, the
generated pions do not have a chance to thermalize. Thus, it is
expected\cite{stephanov,zhuang2} that the resulting pion spectrum
will have a nonthermal enhancement at low momentum.

The dependence of non-strangle hadron masses on temperature $T$
and baryon density $n$ can be well parameterized as\cite{ko}
\begin{equation}
\label{mass} {m(T,n)\over m(0,0)} = \left(1-\left(T\over
T_c\right)^{1/3}\right)\left(1-0.2{n\over n_0} \right)\ ,
\end{equation}
where $T_c$ is the critical temperature of chiral phase
transition, and $n_0$ the baryon density of normal nuclear matter.
From this parameterization, the temperature and density effects
are quite different: The temperature dependence is very weak at
low temperatures ($T/T_c < 1/2$), while the hadron masses decrease
linearly in density in the whole density region. This difference
between temperature and density effects is reflected in the pion
spectra in high energy nuclear collisions. For ultra-relativistic
heavy ion collisions where the central region is almost baryon
free and the temperature effect is dominant, the sigma and pion
masses are nearly constants when the temperature of the system is
much below the critical one. Therefore, the pions produced from
nonthermal sigma decay after freeze-out will have almost the same
momentum $p_0 = \sqrt{{m_\sigma^2(0,0)\over 4}-m_\pi^2(0,0)}$ in
the vacuum. Aside from an overall enhancement of pions due to the
thermal sigma decay, there will be also a peak in the pion spectra
due to the nonthermal sigma decay. This peak will disappear partly
when the density effect can not be neglected, since in this case
the sigma mass varies significantly even after the freeze-out.
Therefore, the sharp peak in the spectrum will be replaced by a
smooth low-momentum pion enhancement.

We now illustrate the above statement by using a simple statistic
model. The measured final-state pions include the direct pions
emitted from the fireball and the pions from sigma decay,
\begin{equation}
\label{total}
N = N_{dir} + N_{dec}\ .
\end{equation}
$N_{dir}$ is determined at freeze-out, and $N_{dec}$ is twice the
number of sigmas at chiral phase transition where sigma has
minimum mass and is then most numerous,
\begin{eqnarray}
\label{dirdec}
&& N_{dir} = V_f n_\pi(T_f)\ ,\nonumber\\
&& N_{dec} = 2 N_\sigma (T_c) = 2 V_c n_\sigma(T_c)\ ,
\end{eqnarray}
where $T_f$ and $V_f$ are the temperature and volume of the system
at freeze-out, $T_c$ and $V_c$ the temperature and volume at
chiral phase transition, $n_\pi(T_f)$ and $n_\sigma(T_c)$ the pion
and sigma number densities at $T_f$ and $T_c$,
\begin{eqnarray}
\label{npns}
&& n_\pi(T_f) = \int{d^3{\bf p}\over (2\pi)^3} {3\over e^{\epsilon_\pi/T_f}-1}\ ,\nonumber\\
&& n_\sigma(T_c) =\int{d^3{\bf p}\over (2\pi)^3} {1\over
e^{\epsilon_\sigma/T_c}-1}\ ,
\end{eqnarray}
with the pion and sigma energies $\epsilon_\pi (T_f)=
\sqrt{m_\pi^2(T_f)+p^2}$ and $\epsilon_\sigma(T_c) =
\sqrt{m_\sigma^2(T_c)+p^2}$. In central region of relativistic
heavy-ion collisions where there is almost no net baryon and one
considers the temperature effect only, the Bjorken's scaling
hydrodynamics\cite{bjorken} is often used to describe the
space-time evolution of the produced hot system. In this scenario
the temperature $T_f$, volume $V_f$ and proper time $\tau_f$ at
freeze-out and the corresponding quantities $T_c, V_c$ and
$\tau_c$ at chiral phase transition have simple relations
\begin{eqnarray}
\label{bjorken}
&& T_f=T_c\left({\tau_c\over \tau_f}\right)^{1\over 3}\ ,\nonumber\\
&& V_f=V_c{\tau_f\over \tau_c}\ .
\end{eqnarray}

For central $A-A$ collisions with impact parameter $b=0$, the
number of participant nucleons can be taken as $\langle
N_{part}\rangle = 2A$ and the volume at freeze-out can be assumed
to be proportional to $A$, $V_f = A v_f $. The pseudorapidity
density of charged pions per participant pair in the mid-rapidity
range $|\eta|<1$ is then written as,
\begin{eqnarray}
&& \label{density} {dN\over d\eta}\Big|_{\Delta\eta}/0.5\langle
   N_{part}\rangle = {1\over 2}{2\over 3}v_f\times\nonumber\\
&& \ \ \ \ \ \left(1+2\left({T_f\over
T_c}\right)^3{n_\sigma(T_c)\over
   n_\pi(T_f)}\right)n_\pi(T_f) \Big|_{\Delta\theta}\ ,
\end{eqnarray}
where $n_\pi(T_f)\Big|_{\Delta\theta}$ is the pion number density
in the angle interval $2\arctan e^{-1} <\theta < 2\arctan e$,
\begin{equation}
\label{theta12}
n_\pi(T_f)\Big|_{\Delta\theta} = \int_{\Delta\theta} {d^3{\bf p}\over (2\pi)^3}
{3\over e^{\epsilon_\pi/T_f}-1}\ ,
\end{equation}
and the factor of $2/3$ is the charged to total pion ratio.

The pseudorapidity density (\ref{density}) depends on the volume
parameter $v_f$ and the two temperature scales $T_f$ and $T_d$
which characterize the collective effect of the system at
freeze-out and at chiral transition. The volume parameter $v_f$ is
centrally colliding energy dependent, it can be fixed by
experimental data. From the lattice simulations of QCD which make
measurements of the Polyakov loop and the scalar quark density,
the deconfinement and chiral transitions coincide at a temperature
of about $T_c = 170\ MeV$\cite{lattice}. In the QCD phase
diagram\cite{heinz} the thermal freeze-out temperature in high
energy nuclear collisions is about $T_f = 120\ MeV$. As is well
known, the pion mass is approximately temperature independent in
the whole chiral breaking phase, we can take $m_\pi(T_f) =
m_\pi(0) = 140\ MeV$ at freeze-out and $m_\sigma(T_c) =
2m_\pi(T_c) =2m_\pi(0)$ at chiral phase transition. With the given
$T_f$ and $T_c$, the ratio of the pions produced by sigma decay to
the direct pions is
\begin{equation}
\label{ratio} r = 2\left({T_f\over T_c}\right)^3
{n_\sigma(T_c)\over n_\pi(T_f)} = 52\%\ ,
\end{equation}
this shows a significant influence of the chiral phase transition
on the final state pions in relativistic heavy-ion collisions.

In above study of pion multiplicity we did not distinguish the
nonthermal sigma decay from thermal sigma decay. As we discussed
qualitatively, the thermal and nonthermal pions have very
different momentum spectra. The pions produced by sigma decay in
the temperature region $T_f < T < T_c$ have time to thermalize
before freeze-out, they lead to an enhancement of thermal pions,
while in the interval $T<T_f$ the produced pions do not get a
chance to thermalize, they have almost the same momentum $p_0 =
\sqrt{{m_\sigma^2(0,0)\over 4}-m_\pi^2(0,0)}$ in the rest frame of
sigma due to the weak temperature dependence of non-strangle
hadron masses at low temperatures. Therefore, if we take a narrow
momentum window around $p_0$, only a small part of the thermal
pions including the direct pions and thermal pions from sigma
decay distribute in the window, but all the nonthermal pions are
deposited it. The contribution from sigma decay to final state
pions is therefore greatly amplified in the window. This strong
pion enhancement can be taken as a signature of chiral phase
transition in relativistic heavy-ion collisions.

Considering a symmetric window $\Delta p$ around $p_0$ and taking
the thermal fraction of the pions produced by sigma decay to be
$\beta$, the pseudorapidity density in the mid-rapidity region and
in the momentum window is
\begin{eqnarray}
\label{density2}
&& {{dN\over d\eta}\Big|_{\Delta\eta,\Delta
   p}\over 0.5\langle N_{part}\rangle} = {1\over 2}
   {2\over 3}v_f\times\nonumber\\
&& \ \ \ \ \ \left[\left(1+2\beta\left({T_f\over
   T_c}\right)^3{n_\sigma(T_c)\over n_\pi(T_f)}\right)
   n_\pi(T_f)\Big|_{\Delta\theta,\Delta p}\right.\nonumber\\
&& \left. \ \ \ \ \ \ \ +2(1-\beta)\left({T_f\over
   T_c}\right)^3
   {n_\sigma(T_c)\over n_\pi(T_f)}n_\pi(T_f) \Big|_{\Delta\theta}
\right]\ ,
\end{eqnarray}
with the pion number density
\begin{equation}
\label{window}
n_\pi(T_f)\Big|_{\Delta\theta,\Delta p} = \int_{\Delta\theta,\Delta p} {d^3{\bf p}\over
(2\pi)^3} {3\over e^{\epsilon_\pi/T_f}-1}
\end{equation}
in the window.

From (\ref{density2}) the pion enhancement factor due to thermal
and nontheral sigma decay in the window is
\begin{eqnarray}
\label{rate}
r_{\Delta p} &=& 2\beta\left({T_f\over T_c}\right)^3
   {n_\sigma(T_c)\over n_\pi(T_f)}+\nonumber\\
&& 2(1-\beta) \left({T_f\over T_c}\right)^3
   {n_\sigma(T_c)\over
   n_\pi(T_f)}{n_\pi(T_f) \Big|_{\Delta\theta}\over
   n_\pi(T_f)\Big|_{\Delta\theta,\Delta p}}\ ,
\end{eqnarray}
it depends on the thermal fraction $\beta$ and the momentum window
width $\Delta p$.

\begin{figure}[ht]
\hspace{+0cm} \centerline{\epsfxsize=6cm\epsffile{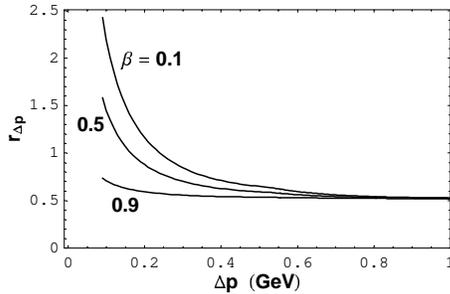}}
\caption{\it The window width dependence of the enhancement factor
(\ref{rate}) for $\beta = 0.1, 0.5$ and $0.9$. } \label{fig1}
\end{figure}

\begin{figure}[ht]
\hspace{+0cm} \centerline{\epsfxsize=6cm\epsffile{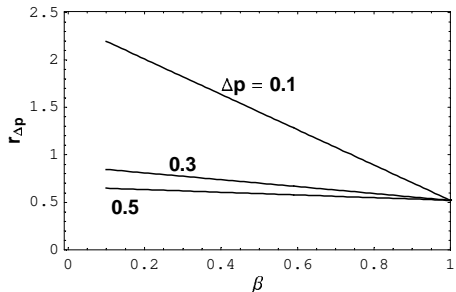}}
\caption{\it The thermal fraction dependence of the enhancement
factor (\ref{rate}) for $\Delta p = 0.1, 0.3$ and $0.5$ GeV. }
\label{fig2}
\end{figure}

In our numerical calculation we take the sigma mass in the vacuum
$m_\sigma (0,0) = 600\ MeV$, the corresponding momentum of the
nonthermal pions is then $p_0 = 265\ MeV$. For a symmetric
momentum window around $p_0$ the enhancement factor $r_{\Delta p}$
is shown in Fig.(\ref{fig1}) as a function of the window width for
three values of the thermal fraction $\beta$ and in
Fig.(\ref{fig2}) as a function of the thermal fraction $\beta$ for
three values of the window width $r_{\Delta p}$. The enhancement
drops down rapidly with increasing window width and becomes
saturated when the width is larger than about $0.4 $ GeV. It is
inversely proportional to the thermal fraction. In the limit of
wide window or large thermal fraction, the nonthermal sigma decay
can be neglected and the enhancement in the window approaches the
total enhancement (\ref{ratio}). It is clear to see that the pion
number in a small momentum window increases dramatically due to
nonthermal sigma decay. Even for $\beta = 0.9$, namely only $10\%$
of the sigmas decay after the freeze-out of the system, the
enhancement in a narrow window is still significant.

%%%%%%%%%%%%%%%%%%%%%%%%%%%%%%%%%%%%%%%%%%%%%%%%%%%%%%%%%%%%%%%%%%%%%%%%
%%%%%%%%%%%%%%%%%%%%%%%%%%%%%%%%%%%%%%%%%%%%%%%%%%%%%%%%%%%%%%%%%%%%%%%%

%%%%%%%%%%%%%%%%%%%%%%%%%%%%%%%%%%%%%%%%%%%%%%%%%%%%%%%%%%%%%%%%%%%%%%%%
%%%%%%%%%%%%%%%%%%%%%%%%%%%%%%%%%%%%%%%%%%%%%%%%%%%%%%%%%%%%%%%%%%%%%%%%

In summary, we have investigated the effect of chiral phase
transition on final state pion spectra. The sigma enhancement at
chiral phase transition leads to a significant enhancement of
pions due to the sigma decay in  chiral symmetry breaking phase.
Especially, the sigma decay after the freeze-out of the system
results in a crucial enhancement in a small momentum window around
$p_0 = \sqrt{{m_\sigma^2\over 4}-m_\pi^2}\sim 260\ MeV$. This
anomalous enhancement comes from the temperature effect which does
not affect the sigma mass at low temperatures. For density effect,
the sigma mass drops down rapidly even at low densities, the pions
from sigma decay after the freeze-out distribute also in a wide
momentum region, then the pion enhancement becomes weak in the
window. Therefore, in central region of relativistic heavy-ion
collisions where there is almost no density effect, the pion
enhancement in a small window around the nonthermal momentum $p_0$
may be considered as a signature of the chiral phase transition.
Although the sigma width and the sigma momentum which are not
taken into account in our estimation will smooth the sharp peak
around $p_0$, we expect that the dynamics of chiral phase
transition (here via temperature dependent masses) plays still an
essential role in the pion spectra.

{\bf acknowledgments}: This work was supported in part by the NSFC
under contract numbers 19925519 and 10135030, and by the Major
State Basic Research Development Program under contract number
G2000077407.


\begin{thebibliography}{99}
\bibitem{qm01}    {For instance, see Proceedings of QM'01,
                   Nucl. Phys. {\bf A698}(2002).}

\bibitem{hwa}     {For instance, see Quark-Gluon Plasma, ed. R. C. Hwa
                   (World Scientific, 1990).}

\bibitem{song}    {Chungsik Song, Volker Koch,
                   Phys. Lett. {\bf B404}(1997)1.}

\bibitem{volkov}  {M.K.Volkov, E.A.Kuraev, D.Blaschke, G.R\"opke, S.Schmidt,
                   Phys. Lett. {\bf B424}(1998)235.}

\bibitem{weldon}  {H.Arthur Weldon,
                   Phys. Lett. {\bf B274}(1992)133.}

\bibitem{chiku}   {S.Chiku and T.Hatsuda,
                   Phys. Rev. {\bf D57}(1998)R6.}

\bibitem{rho}     {M.Rho,
                   Proceedings of the International Workshop XXIII on Gross properties
                   of Nuclei and Nuclear Excitations, Hirschegg, Austria, January, 16-21,
                   1995.}

\bibitem{zhuang}  {P.Zhuang, J.H\"ufner, and S.P.Klevansky,
                   Nucl. Phys. {\bf A576}(1994)525.}

\bibitem{rapp}    {R.Rapp, T.Sch\"afer, E.Shuryak and M.Velkovsky,
                   Phys. Rev. Lett.{\bf 81}(1998)53;
                   M.Stephanov, K.Rajagopal and E.Shuryak,
                   Phys. Rev. Lett. {\bf 81}(1998)4816;
                   K.Rajagopal, hep-ph/9808348;
                   J.Berges and K.Rajagopal, Nucl. Phys. {\bf B538}(1999)215.}

\bibitem{stephanov}{M.Stephanov, K.Rajagopal and E.Shuryak,
                    Phys. Rev. {\bf D60}(1999)114028.}

\bibitem{zhuang2} {P.Zhuang and Z.Yang, Phys. Rev. {\bf D63}(2001)016004.}

\bibitem{ko}      {C.M.Ko, Z.G.Wu, L.H.Xia and G.E.Brown,
                   Phys. Rev. Lett.{\bf 66}(1991)2577; Phys. Rev. {\bf
                   C43}(1991)1881;
                   P.Braun-Munzinger, J.Stachel,
                   Nucl. Phys. {\bf A606}(1996)320.}

\bibitem{bjorken}  {J.D.Bjorken,
                    Phys. Rev. {\bf D27}(1983)140.}

\bibitem{lattice} {F.Karsch,
                   Nucl. Phys. {\bf B}(Proc. Suppl.)83-84(2000)14.}

\bibitem{heinz}   {U.Heinz,
                   Nucl. Phys. {\bf A685} (2001) 414.}

\end{thebibliography}
\end{document}